\documentclass[12pt,preprint]{aastex}

\newcommand{\sollum} {\hbox{$L_{\odot}$}}

\def\deg{\ifmmode^\circ\else$^\circ$\fi}

\shorttitle{GBT Maser Discoveries}
\shortauthors{Braatz et al.}

\begin{document}

\title{A GBT Search for Water Masers in Nearby AGNs}

\author{J. A. Braatz}
\affil{National Radio Astronomy Observatory\altaffilmark{1}, 520 Edgemont Rd., 
Charlottesville, VA 22903}
\altaffiltext{1}{The National Radio Astronomy Observatory is a facility of
the National Science Foundation operated under cooperative agreement with
Associated Universities, Inc.}
\author{C. Henkel}
\affil{Max-Planck-Institut f\"{u}r Radioastronomie, Auf dem H\"{u}gel 69,
D-53121 Bonn, Germany}
\author{L. J. Greenhill\altaffilmark{2} and J. M. Moran}
\altaffiltext{2}{Current address: Kavli Institute for Particle Astrophysics
and Cosmology, Stanford Linear Accelerator Center, 2575 Sand Hill Rd., Menlo
Park, CA 94025, USA}
\affil{Harvard-Smithsonian Center for Astrophysics, 60 Garden Street, 
Cambridge, MA 02138}
\author{A. S. Wilson\altaffilmark{3}}
\affil{Department of Astronomy, University of Maryland, College Park, MD 20742}
\altaffiltext{3}{Adjunct Astronomer, Space Telescope Science Institute,
Baltimore, MD 21218}

\begin{abstract}
Using the Green Bank Telescope, we have conducted a survey for 1.3 cm water
maser emission toward the nuclei of nearby active galaxies, the
most sensitive large survey for H$_2$O masers to date.  Among 145 
galaxies observed, maser emission was newly detected in eleven sources and 
confirmed in one other.  Our survey targeted nearby (v $<$ 12,000 km s$^{-1}$),
mainly type 2 AGNs north of $\delta$ = --20$\deg$, and includes a few additional
sources as well.  We find that more than a third of Seyfert 2 galaxies have
strong maser emission, though the detection rate declines beyond
v $\sim$ 5000 km s$^{-1}$ due to sensitivity limits.  Two of the masers
discovered during this survey are found in unexpected hosts: NGC~4151 
(Seyfert~1.5) and NGC~2782 (starburst).  We discuss the possible relations 
between the large X-ray column to NGC~4151 and a possible hidden AGN in 
NGC~2782 to the detected masers.
Four of the masers discovered here, NGC~591, NGC~4388, NGC~5728 and NGC~6323,
have high-velocity lines symmetrically spaced about the systemic velocity, a
likely signature of
molecular gas in a nuclear accretion disk.  The maser source in NGC~6323, 
in particular, reveals the classic spectrum of a ``disk maser'' represented by 
three distinct groups of Doppler components.  Future single-dish and VLBI
observations
of these four galaxies could provide a measurement of the distance to each
galaxy, and of the Hubble constant, independent of standard candle
calibrations.
\end{abstract}

\keywords{galaxies: active --- galaxies: nuclei --- galaxies: Seyfert ---
ISM: molecules --- masers --- radio lines: galaxies}

\section{Introduction}

Water vapor masers form in warm (300 -- 1000 K), dense ($10^{7} - 10^{11}$
cm$^{-3}$) gas and are common in the envelopes of late-type stars and in
star-forming regions.  They have been detected throughout the Milky Way and in 
many external galaxies as well.  Some extragalactic water masers are
simply analogues to the masers seen in star-forming clouds in our own
galaxy.  Others are found in more exotic locations, especially the
accretion disks and gaseous outflows associated with active galactic nuclei
(AGN).  Owing to their large apparent isotropic luminosity, and following the 
terminology applied to strong extragalactic OH masers, those H$_2$O masers 
detected in active galactic nuclei have been called ``megamasers''.
Review articles on the topic of water megamasers have been written recently 
by Maloney (2002), Greenhill (2002) and Henkel and Braatz (2003).

At 1.3 cm wavelength, water masers can be imaged with mas resolution using
VLBI.  Maps of masers in the nearby Sy~2/LINER nucleus of 
NGC~4258 (Greenhill et al. 1995b; Miyoshi et al. 1995; Herrnstein et al. 1999)
showed that they arise in a thin, warped, edge-on disk at a 
galactocentric radius of 0.14 -- 0.28 pc.  Maser components near the systemic 
velocity form in clouds on the near side of the disk, and ``satellite lines''
with velocities up to $\pm$ 1100 km s$^{-1}$ (Humphreys et al. 2003) from 
systemic form in gas located at points viewed tangentially to its orbital path.
The maser disk in NGC~4258 follows a
Keplerian rotation curve and implies a central mass of 3.9 $\times$
10$^{7}$ M$_\odot$ (Herrnstein et al. 1999).  The satellite lines also
reveal the angular size of the disk as well as its shape (i.e. the warp).

The velocities of maser lines near the systemic velocity in NGC~4258 drift
redward through a $\sim$ 100 km s$^{-1}$ window at a rate of about
9 km s$^{-1}$ yr$^{-1}$ (Haschick, Baan \& Peng 1994; Greenhill et al. 1995a).
The drift reveals the centripetal acceleration of gas in the disk as
it moves across our line of sight to the central core.  
Herrnstein et al. (1999) developed a model of the maser disk based on
its VLBI structure, and combined their derived disk parameters with the
measured acceleration to calculate the linear size of the disk.
Comparing the linear size of the disk with its angular size, they
determined a distance to the galaxy independent of the usual use of standard 
candles.  The authors also used the independent technique of proper motions
to confirm the result.  Establishing a set of distances to other galaxies 
using these techniques would provide a new foundation to the extragalactic 
distance ladder uncoupled from other methods that have historically involved 
large systematic uncertainties.

There has been a resurgent interest in surveys for extragalactic water maser
sources in recent years, inspired by both the above issues and new 
instrumentation.  Programs by Hagiwara, Diamond and Miyoshi (2002), 
Henkel et al. (2002), Greenhill et al (2003), Hagiwara et al. (2003), 
Kondratko et al. (2003), and Peck et al. (2003), 
among others, have increased the number of known extragalactic
H$_2$O masers to $\sim$ 50.  Still, only a few of these masers have the
signature of a nuclear disk, and more examples are needed.

In this paper we discuss the results of a new survey for H$_2$O maser emission
from a sample of nearby (v $<$ 12,000 km s$^{-1}$) galaxies, nearly all AGNs.  
With the goal of maximizing the detection rate and searching for disk maser 
candidates, we concentrated on type 2 Seyferts and LINERs, which are known
to be the most likely hosts of detectable masers (Braatz, Wilson \& Henkel
1997).  The candidates were primarily selected from the CfA Seyfert Catalog
(Huchra, available at http://cfa-www.harvard.edu/$\sim$huchra), a collection
of AGNs drawn from CfA redshift surveys and identifications of X-ray and
infrared sources.  Eight galaxies in our survey are not in this catalog but 
were of interest because of strong infrared luminosity or known nonstellar 
nuclear activity.  These are IC~694, IC~4553, NGC~3660, NGC~4013, NGC~5635, 
NGC~5953, NGC~6211, and UGC~3995A.  

Each of the galaxies observed during the 
present survey has been observed in at least one previous survey for water 
vapor maser emission, e.g. by Braatz, Wilson \& Henkel (1996), Greenhill 
et al. (1997) or Taylor et al. (2002).  The 
motivation to reobserve these sources comes primarily from recent advances in 
K-band sensitivity, especially due to the availability of the Robert C. Byrd
Green Bank Telescope (GBT).  In addition, the availability of large bandwidth
($>$ 200 MHz) spectrometer modes now makes it possible to search efficiently
for high-velocity lines.  Finally, the intrinsic variability of H$_2$O masers 
makes it worthwhile to reobserve sources even if they have been previously 
undetected.  Galaxies already known to host H$_2$O masers were not observed 
during this survey, but are being studied as part of other programs, e.g. 
Braatz et al. (2003).

\section{Observations}

Observations were made with the GBT during several sessions between 
2003 March 4 and 2004 February 28.  This period marked a steady
improvement in the GBT's high frequency observing capabilities, particularly
with regard to baseline stability, pointing accuracy, focus tracking, and
sensitivity.  We used the 18 -- 22 GHz K-band receiver, which has two beams at
a fixed separation of 3$\arcmin$ in azimuth.  The GBT beam width is $\sim$
36$\arcsec$ at 22 GHz, and pointing uncertainties were 3$\arcsec$ -- 8$\arcsec$.
The telescope was nodded between two positions on the sky such that the source
was always in one of the two beams during integration.  Prior to 2003 June 4,
we observed with the electronic beam switch in the receiver cycling at a rate 
of 1 Hz, and used a nod cycle of 5 minutes.  After that date, we observed
with no electronic beam switching, and we shortened the nod cycle to 2 
minutes.  The latter configuration provided equally good baselines, a simpler 
data format, and better system reliability since a sometimes faulty electronic
switch was bypassed.

The spectrometer was configured with two 200 MHz bandpasses, one centered
on the systemic velocity of the galaxy and the second redshifted by 180 MHz.
Channel spacing in the spectra corresponds to 0.33 km s$^{-1}$.  The zenith 
system temperature was between 35K and 80K.  Atmospheric opacity was 
either measured with tipping scans or estimated from system temperature and
weather data, and ranged from 0.04 to 0.12 at the zenith.  Data were reduced 
using AIPS++.  When baseline stability permitted, we smoothed the reference 
spectra prior to calibration using a 16 channel boxcar function, to improve 
sensitivity.  In most cases we subtracted a polynomial (order between 3 and 6) 
from the 200 MHz spectrum to remove the baseline shape.  For some galaxies we 
subtracted a baseline structure derived from the lowest frequency Fourier 
components of the spectrum.  The 1$\sigma$ r.m.s. sensitivity of the survey 
ranges between 3 and 6 mJy per 24.4 kHz ($\sim$0.33 km s$^{-1}$) channel.  
This sensitivity is equivalent to roughly 0.5 $\sollum$ for a 1 km s$^{-1}$ 
line in a galaxy with a recession velocity of 5000 km s$^{-1}$.

Velocities throughout this paper follow the optical definition of Doppler 
shift and are calculated with respect to the Local Standard of Rest (LSR).
Galaxy recession velocities are derived either from 21 cm HI observations
or optical observations.  Galaxy distances are calculated using H$_0$ = 75 
km s$^{-1}$ Mpc$^{-1}$.

\section{Results and Discussion}

We observed 145 nearby AGNs during the survey, and detected masers in 12
galaxies (Table 1), giving a detection rate of 8.3\%.  Without our knowledge 
at the time of the observation, one of these masers (in Mrk 1066) had been 
detected at Effelsberg by Peck et al (2003); the other 11 are new discoveries.
Maser emission has been detected in each of the 12 galaxies with the GBT on 
at least two epochs, and
the best spectrum for each one is shown in Figure 1.  A complete presentation
of data and an analysis of the variability in these sources will be presented 
in a later publication.  The total number of galaxies, excluding the Milky Way,
with detected H$_2$O masers is now 63.  Table 1 lists the galaxies
detected during this survey and Table 2 lists those observed but not detected.
Each galaxy was observed toward its nucleus.  No maser emission was detected 
in any of the galaxies in the second bandpass, shifted by 180 MHz.  A more 
comprehensive list of galaxies observed for H$_2$O maser emission, including 
this survey and others, is available from http://www.nrao.edu/$\sim$jbraatz, 
and a list of known extragalactic maser systems is available at the site as 
well.

The galaxies observed during this survey do not form a complete sample but 
the results can still be used to improve our knowledge of the incidence
of detectable masers in AGNs.  Among galaxies listed as Seyfert 2 type in the 
CfA AGN list (see Sect.\,1) and observed with the GBT (either during the 
present survey or during
previous observations of known H$_2$O masers, e.g. Braatz et al. 2003), 
the fraction detected is 17/82 = 21\%,
14/55 = 25\%, and 11/39 = 38\% for galaxies within 12,000 km s$^{-1}$,
7500 km s$^{-1}$, and 5000 km s$^{-1}$, respectively.  Such a decrease of
detection rate with increasing distance reflects the sensitivity limitation of
surveys such as this (see Braatz, Wilson \& Henkel 1997).  In the listed 
distance categories, there were 15, 9, and 2 galaxies, respectively, from the 
CfA list not observed
during this survey.  The LINER sample is less complete than the Seyfert~2
sample, so likewise statistics are less instructive for this class of AGN.
Still we note that the detection rate is 5/42 = 12\%, 5/40 = 13\%, and 4/36 =
11\% for LINERs from the present survey in the same three distance categories.

Two of the sources detected in this survey, NGC~2782 and NGC~4151, are 
included in Carl Seyfert's original list of twelve ``Emission Line Galaxies''
(Seyfert, 1943) that defined the Seyfert type, and it is interesting to note
that five of those twelve have now been detected in H$_2$O maser emission.
However, NGC~2782 and NGC~4151 are not usually classified as type 2 AGNs, 
counter to the trend of most other extragalactic H$_2$O masers.  NGC~2782
has a powerful nuclear starburst but the high luminosity of the water masers
(L$_{iso} = 12 \sollum$) suggests the presence of an AGN.  Kennicutt, Keel \&
Blaha (1989) find that the profiles of the H$\beta$ and [OIII]$\lambda\lambda$
4959, 5007 lines are dramatically different, with [OIII] showing both a 
broader core
and high velocity wings.  They conclude that the high velocity gas in the wings
of [OIII] must have very high excitation, reminiscent of the narrow line
emission in Seyfert galaxies, and that the nucleus may be a ``composite''
(i.e. starburst plus AGN) system (cf. V\'{e}ron, Gon\c{c}alves \& 
V\'{e}ron-Cetty 1997).
Other workers (e.g. Schulz et al. 1998) have, however, interpreted the
high excitation gas in terms of a starburst-driven superwind, so the
evidence for an AGN in NGC~2782 is not compelling.  A hard X-ray spectrum of
this galaxy could settle the issue.

NGC~4151 is a nearby (13.3 Mpc) active galaxy usually classified as a 
Seyfert 1.5 (e.g., Osterbrock \& Koski 1976), but the broad emission lines 
in its optical spectrum are variable (e.g. Sergeev, Pronik \& Sergeeva 2001) 
giving it a Seyfert 2 profile at times.  The galaxy has a large X-ray column 
density toward its nucleus ($\sim$ 10$^{23}$ cm$^{-2}$, Weaver et al. 1994).
The proximity of NGC~4151 and this large X-ray column are conducive to the
detection of water vapor emission.  Braatz et al. (1997) concluded that the
probability of H$_2$O maser emission in Seyfert galaxies increases with
increasing X-ray inferred column density to the nucleus.  The X-ray measured 
column density to the nucleus of NGC~4151 is roughly two orders of magnitude
larger than that toward a typical Seyfert 1, and
is similar to the column found among typical Seyfert 2s
(Turner et al. 1997).  At 0.7 $\sollum$, the isotropic luminosity of the maser 
in NGC 4151 is among the weakest of any detected in an AGN.  The maser emission
is primarily confined to two narrow components (see Figure 1), one at 692.4 
km s$^{-1}$ and the other at 1126.6 km s$^{-1}$.  The widths of the emission 
lines, determined from Gaussian fits, are 1.2 and 1.5 km s$^{-1}$ (full width 
at half maximum), respectively.
NGC 4151 has been searched previously for H$_2$O emission, but not detected.  
The most sensitive previous observation, obtained on 2000 March 20, resulted 
in a 3$\sigma$ detection limit of 9 mJy per 4.2 km s$^{-1}$ channel 
(Taylor et al. 2002).  Our detection shows the lines having peak flux
densities of 36 and 53 mJy (Figure 1), but the lines are narrower than the 
channel spacing used by Taylor et al.  The peak flux densities in our 
spectrum are reduced to $\sim$11 and $\sim$19 mJy after averaging to a channel 
spacing of 4.2 km s$^{-1}$.  That these fluxes are still above the Taylor 
et al.  detection limit may reflect intrinsic variability in the NGC 4151 
maser.

Three sources detected in this survey are associated with merging systems:
NGC~2782, NGC~4922, and NGC~5256.  In NGC 4922 the 
separation of the merging nuclei is 22$\arcsec$ (10 kpc) and each nucleus is 
itself an AGN.  NGC 5256 is similar, with two AGNs in the merger separated by 
11$\arcsec$ (5.9 kpc).  In both systems, an observation was made toward each 
of the nuclei to determine the source of the H$_2$O emission.  In NGC~4922
the observations are consistent with all of the detected maser emission 
originating from component NGC~4922N (the more northern component) and in 
NGC~5256 the observations are consistent with all of the detected maser 
emission originating in NGC~5256S (the more southern component).

The masers detected in NGC 591, NGC 4388, NGC 5728 and NGC 6323 all reveal
high-velocity features approximately symmetrically spaced about the systemic 
velocity of the galaxy, a possible signature of a nuclear disk.  In this
interpretation, the maximum rotation velocities would be $\sim$425, $\sim$400, 
$\sim$250 and $\sim$550 km s$^{-1}$ respectively.
Two of these galaxies, NGC 4388 and NGC 5728, are known to have a well defined,
bi-conically shaped narrow line region (Pogge 1988; Corbin, Baldwin \& Wilson
1988; Wilson et al. 1993), a characteristic feature of a highly inclined, 
optically thick nuclear torus around the source of ionizing radiation.
NGC 6323 has the classic maser spectrum expected from a nuclear disk, with
a distinct cluster of Doppler components near its systemic velocity in addition 
to the high-velocity lines.  At a distance of $\sim$100 Mpc, 14 times more
distant than NGC~4258, NGC~6323 is a strong candidate for detailed VLBI studies 
that could ultimately provide a direct measurement of the Hubble constant.
Because of the faintness of its masers, a global array of telescopes including 
the GBT, Effelsberg, and Goldstone in addition to the VLBA would be essential
to map the disk.

\section{Summary}

We searched 145 nearby active galaxies for 1.3 cm H$_2$O maser emission and 
detected 12 galaxies.  Although each of these galaxies has been observed
but not detected in H$_2$O during previous surveys, we attained a fairly
high detection rate of 8.3\% due to the improved sensitivity of this survey.
Four of the newly discovered masers have spectral profiles consistent with a
disk maser and might become useful subjects of followup observations to map 
the disk, characterize the kinematics, and perhaps make direct measurements of 
the distances to the host galaxies.  Our results indicate that greater than a
third of Seyfert 2 galaxies host water megamasers, though the detection rate
declines for galaxies beyond $\sim$ 5000 km s$^{-1}$ due to sensitivity 
limits.  Although Seyfert 2 galaxies are the most likely class of maser
host, one of the masers discovered in this survey is in the type 1.5 Seyfert
NGC~4151, and another is in the starburst galaxy NGC~2782.  We discuss
possible reasons for luminous water maser emission in these two galaxies.  
Our survey demonstrates that the recent improvements in sensitivity and 
bandwidth warrant more extensive surveys, including type 1 AGNs, starburst 
galaxies, and apparently normal galaxies.

\begin{table}
\caption{H$_2$O Masers Detected During the GBT Survey}
\bigskip
\begin{tabular}{lrrrrrr}
\tableline\tableline
Source & $\alpha_{2000}$ & $\delta_{2000}$ & V$_{LSR}$\tablenotemark{a} & $S_{peak}$\tablenotemark{b} & L$_{iso}$\tablenotemark{c} & Date of Observation \\
 & (\ h\ \ \ m\ \ \ s\ )&( $\ \deg\ \ \ \arcmin\ \ \ \arcsec\ \ $) & (km s$^{-1})$ & (mJy) & ($\sollum$) &  \\
\tableline
NGC~591  & 01 33 31.2 & +35 40 06 &  4554 $\pm$ 9 & 10  & 25  & 31 Jan 2004 \\
Mrk~1066 & 02 59 58.6 & +36 49 14 &  3601 $\pm$ 22 & 48  & 30  & 28 Jan 2004 \\
UGC~3255 & 05 09 50.2 & +07 29 00 &  5674 $\pm$ 59 & 15  & 16  & 6 Mar 2003 \\
Mrk~3    & 06 15 36.3 & +71 02 15 &  4009 $\pm$ 6  & 17  & 11  & 28 Jan 2004 \\
Mrk~78   & 07 42 41.7 & +65 10 37 & 11196 $\pm$ 29 & 28  & 34  & 15 Nov 2003 \\
NGC~2782 & 09 14 05.1 & +40 06 49 &  2560 $\pm$ 5  & 45  & 12  & 14 Jan 2004 \\
NGC~4151 & 12 10 32.6 & +39 24 21 &  1002 $\pm$ 3  & 52  & 0.7 & 28 Feb 2004 \\
NGC~4388 & 12 25 46.7 & +12 39 44 &  2521 $\pm$ 4  & 15  & 12  & 14 Jan 2004 \\
NGC~4922 & 13 01 25.2 & +29 18 50 &  7080 $\pm$ 9  & 36  & 190 & 31 Oct 2003 \\
NGC~5256 & 13 38 17.2 & +48 16 32 &  8365 $\pm$ 13 & 99  & 30  & 4 Nov 2003 \\
NGC~5728 & 14 42 23.9 & --17 15 11 &  2796 $\pm$ 8  & 173 & 88  & 14 Jan 2004 \\
NGC~6323 & 17 13 18.0 & +43 46 56 &  7791 $\pm$ 35 & 41  & 480 & 2 Jun 2003 \\
\tableline
\end{tabular}
\tablenotetext{a}{Recession velocity measured with respect to the local 
standard of rest.  Velocities and uncertainties are from \citet{deV91}, using
HI velocities when available and optical velocities otherwise.  For NGC 6323
the optically measured velocity is from Marzke, Huchra \& Geller (1996).}
\tablenotetext{b}{Peak flux density}
\tablenotetext{c}{Inferred isotropic luminosity of the maser emission.}
\end{table}

\begin{table}
\caption{Galaxies Undetected in H$_2$O Emission}
\bigskip
\begin{tabular}{lllll}
\tableline
0152+0622  & IC 4553  & NGC 334  & NGC 3660  & NGC 5427  \\
0253-1641  & II Zw 101  & NGC 404  & NGC 3898  & NGC 5635  \\
0258-1136  & Mrk 176  & NGC 600  & NGC 3921  & NGC 5674  \\
0335+09  & Mrk 198  & NGC 788  & NGC 3982  & NGC 5675  \\
0354-1855  & Mrk 273  & NGC 1144  & NGC 3998  & NGC 5695  \\
0414+00  & Mrk 298  & NGC 1167  & NGC 4013  & NGC 5851  \\
0445-1741  & Mrk 334  & NGC 1229  & NGC 4036  & NGC 5899  \\
0446-2349  & Mrk 359  & NGC 1358  & NGC 4074  & NGC 5929  \\
07570+2334  & Mrk 372  & NGC 1365  & NGC 4111  & NGC 5953  \\
0816+211  & Mrk 403  & NGC 1409  & NGC 4117  & NGC 6211  \\
0942+09  & Mrk 423  & NGC 1410  & NGC 4192  & NGC 6251  \\
1034+060  & Mrk 461  & NGC 1667  & NGC 4278  & NGC 6500  \\
1116-2909  & Mrk 477  & NGC 1685  & NGC 4303  & NGC 6764  \\
1258-3208  & Mrk 516  & NGC 2110  & NGC 4419  & NGC 6951  \\
1319-162  & Mrk 573  & NGC 2273  & NGC 4450  & NGC 7212  \\
1322+2918  & Mrk 612  & NGC 2377  & NGC 4486  & NGC 7217  \\
1335+39  & Mrk 622  & NGC 2768  & NGC 4501  & NGC 7450  \\
1431-3237  & Mrk 745  & NGC 2841  & NGC 4569  & NGC 7674  \\
1533+14  & Mrk 883  & NGC 2911  & NGC 4579  & NGC 7682  \\
1548-0344  & Mrk 917  & NGC 2992  & NGC 4941  & NGC 7743  \\
2319+09  & Mrk 937  & NGC 3010b  & NGC 5005  & [SP] 55  \\
3C 317  & Mrk 955  & NGC 3031  & NGC 5128  & UGC 3995A  \\
Ark 539  & Mrk 1058  & NGC 3185  & NGC 5135  & UGC 6100  \\
Fair 1140  & Mrk 1073  & NGC 3227  & NGC 5195  & UGC 10567  \\
Fair 1149  & Mrk 1098  & NGC 3362  & NGC 5252  & UGC 12056  \\
IC 614  & Mrk 1239  & NGC 3561  & NGC 5283  &   \\
IC 694  & Mrk 1388  & NGC 3642  & NGC 5371  &   \\
\tableline
\end{tabular}
\end{table}

\acknowledgements
We would like to extend our appreciation to the GBT operators and the Green 
Bank staff for support during this program.  We 
thank Don Wells for helpful suggestions concerning analysis techniques, and Ron 
Maddalena for technical assistance with the GBT.  We also thank Bob Garwood
and Joe McMullin for their significant contributions to the data analysis
system in AIPS++.  This research has made
use of the NASA/IPAC Extragalactic Database (NED) which is operated by the Jet 
Propulsion Laboratory, California Institute of Technology, under contract with
the National Aeronautics and Space Administration.
This research was supported in part by NASA through grant NAG 513065 to the
University of Maryland.

\clearpage
\begin{figure}
\epsscale{.75}
\plotone{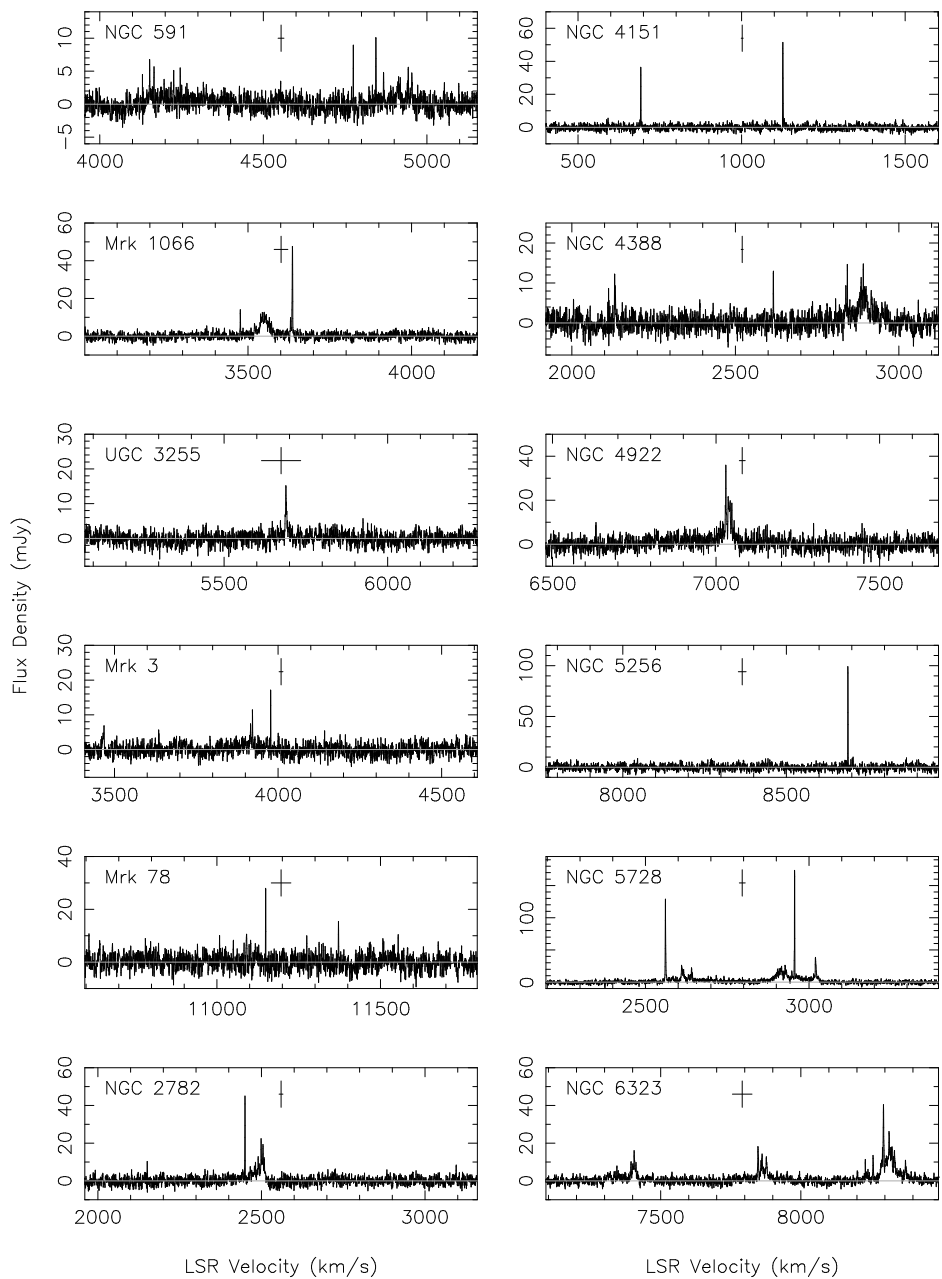}
\figurenum{1}
\caption{Spectra of 1.3 cm water maser emission toward the nuclei
of 12 galaxies.  Each spectrum covers 1200 km s$^{-1}$ except the one toward
NGC~6323, which covers 1400 km s$^{-1}$.  The systemic velocity of each galaxy 
and its uncertainty are marked above each spectrum by a cross.  Velocities 
are measured with respect to the LSR and use the optical definition of
Doppler shift.}
\end{figure}

\end{document}